# Comeback of epitaxial graphene for electronics: large-area growth of bilayer-free graphene on SiC

*Mattias Kruskopf, Davood Momeni Pakdehi, Klaus Pierz, Stefan Wundrack, Rainer Stosch, Thorsten Dziomba, Martin Götz, Jens Baringhaus, Johannes Aprojanz, Christoph Tegenkamp, Jakob Lidzba, Thomas Seyller, Frank Hohls, Franz J. Ahlers and Hans W. Schumacher*

**Abstract**

We present a new fabrication method for epitaxial graphene on SiC which enables the growth of ultra-smooth defect- and bilayer-free graphene sheets with an unprecedented reproducibility, a necessary prerequisite for wafer-scale fabrication of high quality graphene-based electronic devices. The inherent but unfavorable formation of high SiC surface terrace steps during high temperature sublimation growth is suppressed by rapid formation of the graphene buffer layer which stabilizes the SiC surface. The enhanced nucleation is enforced by decomposition of polymer adsorbates which act as a carbon source. With most of the steps well below 0.75 nm pure monolayer graphene without bi-layer inclusions is formed with lateral dimensions only limited by the size of the substrate. This makes the polymer assisted sublimation growth technique the most promising method for commercial wafer scale epitaxial graphene fabrication. The extraordinary electronic quality is evidenced by quantum resistance metrology at 4.2 K with until now unreached precision and high electron mobilities on mm scale devices.

**Main Text**

The success of graphene as a basis for new applications depends crucially on the reliability of the available technologies to fabricate large areas of homogenous high quality graphene layers. Epitaxial growth on metals as well as on SiC substrates is employed with specific benefits and drawbacks. Single graphene layers epitaxially grown on SiC offer a high potential for electronic device applications. They combine excellent properties, e.g. high electron mobilities, with the opportunity for wafer-scale fabrication and direct processing on semi-insulating substrates without the need to transfer the graphene to a suitable substrate (Avouris & Dimitrakopoulos 2012). Some progress has been achieved during the recent years. In particular, high temperature sublimation growth under Ar atmosphere (Virojanadara et al. 2008),(Emtsev et al. 2009) or by confinement control (Heer et al. 2011), (Real et al. 2012) was a breakthrough for synthesizing large-area graphene on SiC substrates. The coverage of graphene bilayers could be reduced from wide stripes formed along the terraces to small micrometer-sized bilayer patches (Virojanadara et al. 2009). Further it was found that beyond pure sublimation growth from SiC graphene formation can be assisted by additional carbon supply from external sources (Al-Temimy et al. 2009; Moreau et al. 2010). In particular, by using propane in a chemical vapor deposition process the versatility of the graphene growth is improved. (Michon et al. 2010; Strupinski et al. 2011)

However, in spite of the progress achieved so far, the growth of high quality graphene in a reproducible manner remains challenging. (Eriksson et al. 2012) The problem of the inherent high step edge formation caused by step bunching of the SiC substrate is not solved, leading to an increased electrical resistance (Ji et al. 2012)(Low et al. 2012) and anisotropic electronic properties (Yakes et al. 2010)(Schumann et al. 2012). Regardless of the fabrication techniques the resulting graphene is not entirely free of bilayer patches which can shortcut electronic structures (Yager et al. 2013)(Chua et al. 2014) (Kruskopf et al. 2015) and severely deteriorate the properties of electronic devices. These drawbacks so far have prevented that SiC sublimation growth is implemented in wafer-scale device fabrication.

Here we describe a new sublimation growth method that leads to the formation of bilayer-free graphene on SiC with exceptionally shallow step heights and a layer size which is only limited by the dimensions of the sample. The core of this method is to form the surface stabilizing buffer layer by an external carbon source before the smooth surface morphology is destroyed by step bunching occurring at high temperature annealing. A simple and effective implementation of this idea is the deposition of a polymer on the SiC substrate which then assists the formation. The process of what we call polymer-assisted sublimation growth (PASG) of graphene on SiC is schematically depicted in Fig. 1 and comprises four steps: (I) deposition of the polymer adsorbates, (II) decomposition into amorphous carbon and nanocrystalline graphite, (III) conversion into buffer layer domains and (IV) closing of the buffer layer and graphene growth by high temperature sublimation growth.

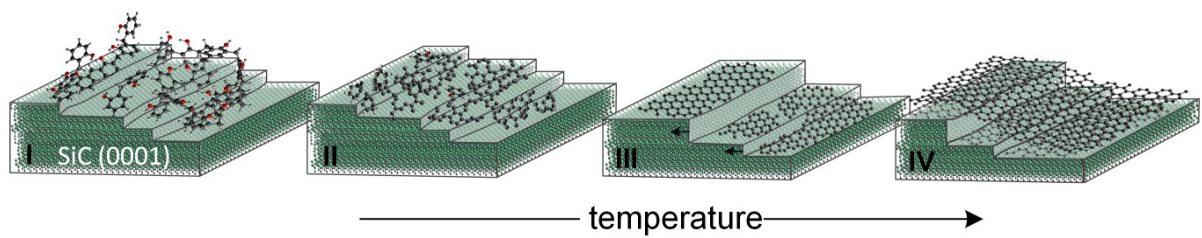

**Figure 1: Schematic of polymer-assisted sublimation growth of graphene on SiC(0001) in four consecutive steps**. **(I)** Deposition of polymer adsorbates. **(II)** Thermal decomposition and crosslinking between ~450 and ~950 °C into disordered carbon networks. **(III)** Conversion into buffer layer domains and silicon sublimation ≥ 1300 °C and initial restructuring of the terraces by retracting step edges **(IV)** Closing of gaps in the buffer layer by sublimation growth and transformation into graphene by growth of a new buffer layer underneath between ~1400 and ~1750 °C.

**Enhanced buffer layer nucleation from graphite nanocrystals**

For PASG a phenolic resin was deposited onto the sample surface by dipping and rinsing (see methods) which results in a homogenous distribution of droplet-like adsorbates with structure heights between 2 and 20 nm, see atomic force microscopy (AFM) image in Fig. 2a. Because of the high thermal stability of the resin the adsorbate remains up to about 300 °C during vacuum annealing. When the annealing temperature exceeds $T_a$ = 450 °C, a disordered carbon network including $sp^2$-hybridized hexagonal carbon rings appears due to crosslinking of the phenolic resin. This is indicated by the two broad Raman bands known as the D- and G- peaks, emerging at around 1360 and 1590 cm$^{-1}$, respectively (Fig. 2k). (Ferrari & Robertson 2000) (Ko et al. 2000) In addition, the X-ray photoemission spectrum (XPS) (Fig. 2l bottom) shows an emerging band upon annealing at 450 °C which can be fitted by two Gaussian curves indicating the presence of two components ($A_1$ and $A_2$) at binding energies of 1.80 eV and 2.70 eV with respect to the SiC peak. To compensate for the shift due to charging effects of the semi-insulating substrates the spectra were aligned in Fig. 2l to the absolute peak position of the SiC bulk component at 283.7 eV (Emtsev et al. 2008). The separation of (0.9 ± 0.1) eV between $A_1$ and $A_2$ is in good agreement with that of the sp² and sp³ hybrid forms of carbon. (Díaz et al. 1996) The appearance of both bonding types indicates the amorphous nature of the carbon network. A portion of ~17% $sp^3$ bonds was estimated from the intensity ratio of the two XPS-components comparable with ~10% derived from Raman-spectroscopy, see supplementary data (Díaz et al. 1996; Ferrari & Robertson 2000).

The AFM image in Fig. 2b shows that after annealing at $T_a$ = 950 °C the high structures of the adsorbate clearly observed in Fig.2a are converted into a flat (wafer manufacturing related) surface with single SiC crystal planes and corresponding step height of 0.25 nm. The low energy electron diffraction (LEED) image of this sample shows a (1×1) pattern of an unreconstructed SiC surface (Fig. 2j). At this stage (Fig. 2b) the carbon network appears planarized but remains detectable by a high density of shallow (≤ 0.5 nm) structures which increase the surface roughness and are attributed to nanocrystalline graphite clusters (Ferrari & Robertson 2000). From the intensity ratio of the Raman bands I(D)/I(G) ~1 a lateral crystallite size of the about 5 nm is estimated, see supplementary data (Tuinstra & Koenig 1970; Ferrari & Robertson 2000). This is in good agreement with the XPS spectrum (Fig. 2l middle) which shows a single peak G1 indicating the remaining $sp^2$ bonds of the nanocrystalline graphite clusters.

The conversion of the carbon clusters into the buffer layer starts at $T_a \approx 1300$ °C. At this temperature a weak LEED pattern of a ($6\sqrt{3} \times 6\sqrt{3}$) surface reconstruction is observed which at 1400 °C turns into a clear pattern shown in Fig. 2i. This indicates an increasing trend towards the formation of connected buffer layer domains, also documented by the change of the vibrational modes in the Raman spectra (Fig. 2k). In the temperature range 1400 - 1500 °C an additional set of Raman bands (around 1370, 1492, 1543 and 1595 cm$^{-1}$) occurs which can be assigned to the vibrational density of states of the ($6\sqrt{3} \times 6\sqrt{3}$) surface reconstruction. (Fromm et al. 2013) This is also confirmed by the corresponding XPS spectrum (Fig. 2l top) showing the typical S1 and S2 components at (1.36 ± 0.15) eV and (2.00 ± 0.19) eV with respect to the SiC-bulk peak position. (Emtsev et al. 2008) The absence of additional components clearly proves that the amorphous carbon is completely incorporated into the buffer layer.

The evolution of the buffer layer and the terrace structure is visualized by AFM measurements taken from a series of three samples that were annealed for 5 min at $T_a$ = 1300, 1400 and 1500 °C in Ar atmosphere, respectively (Fig. 2c-h). The topographic AFM image (Fig. 2c) shows that at 1300 °C regular facets with step heights of 0.25 nm and 0.5 nm evolve. The dense layer of nanometer sized bright spots observed in the AFM phase contrast image (Fig. 2d) is attributed to the first buffer layer domains detected by LEED. The formation of these small domains at 1300 °C from the graphite nanocrystals is crucial since they act as nuclei for diffusing carbon atoms released by the restructuring terraces. The high density of preferred nuclei reduce the mass transport between neighboring terrace edges and thus suppress the major mechanisms that leads to giant step bunching (Jeong & Weeks 1998). A significantly higher annealing temperature of about 1400 °C is needed to initiate decomposition of the shallow 0.25 nm terraces (Fig. 2e) and 1500 °C to reach the next stable step configuration with heights of 0.75 nm (Fig. 2g). The reason for the relatively slow step velocity of the retracting crystal planes is the high density of buffer layer nuclei which stabilize the surface by surface free energy minimization due to covalent bonds to the substrate (Riedl et al. 2010). As a result of the enhanced buffer layer nucleation, domains merge and are collected along the upper side of the retracting terrace edges.

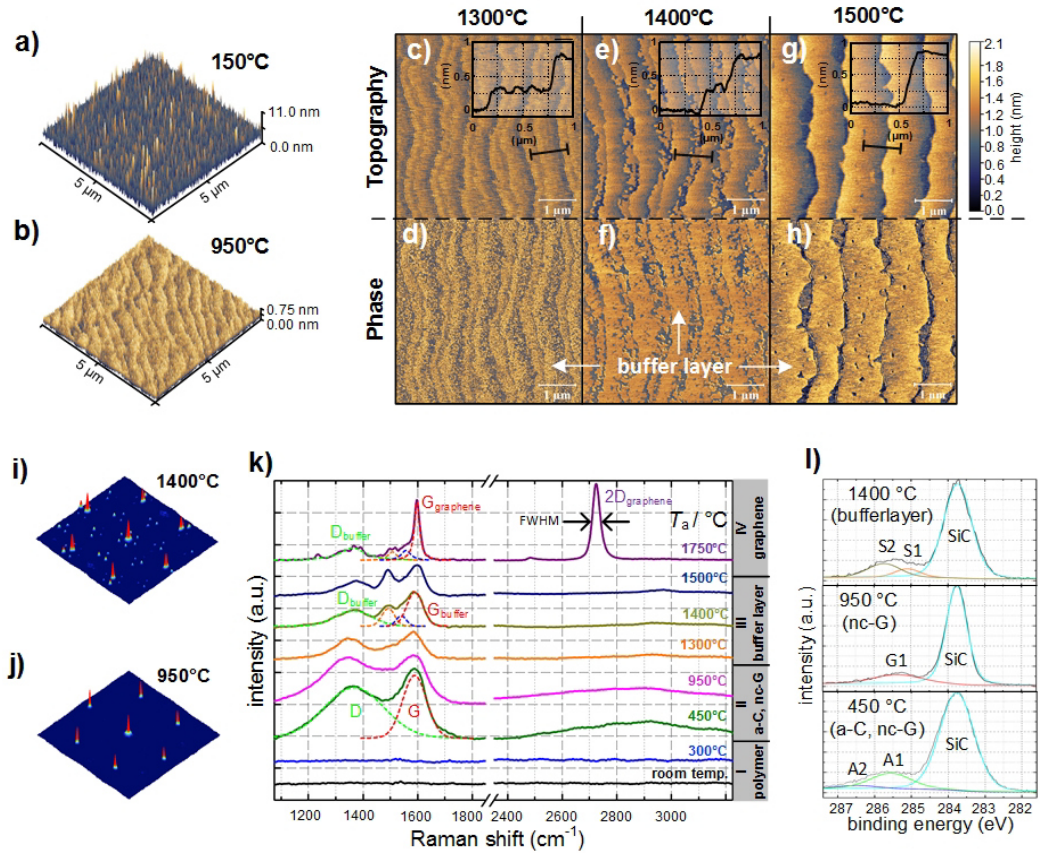

**Figure 2: Formation of the graphene buffer layer from polymer adsorbates. a,** AFM image of the nanometer-size polymer droplets on the SiC surface and **b,** reduced structure heights after vacuum annealing at 950 °C due to conversion into carbon nanocrystals on the shallowly stepped SiC surface. **c,e,g,** AFM topography of three polymer pretreated samples after annealing at 1300, 1400 and 1500 °C. Insets show the height profile along the indicated line. **d,f,h,** Corresponding phase images document the increasing buffer layer coverage indicated by the bright contrast. **d,** From small buffer layer islands at $T_a$ = 1300 °C towards (**f**) large connected domains and (**h**) to almost full coverage of the terraces after annealing at $T_a$ = 1500 °C. **j,** (1×1) LEED pattern of the SiC surface after annealing at 950 °C indicates an unreconstructed SiC surface. **i,** For higher annealing temperatures the LEED pattern develops into a clear ($6\sqrt{3} \times 6\sqrt{3}$) buffer layer reconstruction pattern at 1400 °C. **k,** Room-temperature Raman spectra of 8 polymer pretreated samples measured after annealing at the indicated temperatures $T_a$. The spectrum of the SiC substrate spectrum was subtracted. The four consecutive formation steps shown in Figure 1 are indicated at the right hand side. **l,** XPS C1s core level spectrum of three samples of this series. Bottom: after vacuum annealing at 450 °C two components of disordered carbon, A1 and A2, are observed and the SiC bulk peak. From the fit curves A1 and A2 the fractions of sp² and sp³ hybridized carbon is estimated. Middle: after argon annealing at 950 °C only peak G1 remains which is attributed to sp² bonds of nano-crystalline graphite. Top: after argon annealing at 1400 °C only the typical S1 and S2 buffer layer components and the SiC peak are detected. The XPS spectra are aligned to the same SiC bulk peak energy.

**Bilayer-free graphene growth on ultra-shallow steps**

Monolayer graphene growth was performed on substrates with a small miscut angle (~0.05 °) by pre-annealing at $T_a$ = 1300 °C followed by heating the sample to 1750 °C. The heating rate (~400 K/min) is fast compared to the slow retraction velocity of the steps and should suppress step bunching and the formation of 0.75 nm steps in the temperature range around 1500 °C. Indeed, the AFM topography (Fig. 3a) shows a very regular surface morphology of the resulting graphene layer where about 85 % of the step heights correspond to a sequence of 0.25 nm or 0.5 nm. These heights describe one or two SiC crystal planes, respectively. The absence of heights > 0.75 nm is unique to this method. The step height histogram (Fig. 3b) reveals that for buffer layer formation at higher annealing temperatures of 1400 °C (blue) or 1500 °C (red) the fraction of steps with heights < 0.75 nm decreases to 54 % and 40 %, respectively. This correlation indicates that the final step height found after graphene formation can be traced back to the one which was predefined during the annealing step. The stabilization of the SiC surface which leads to the reduced step velocity is due to the rapidly preformed buffer layer. (Hannon & Tromp 2008; Kruskopf et al. 2015). This prevents restructuring of the SiC surface and a large number of the preformed shallow single and double SiC bilayer steps remain during the high temperature graphitization step at 1750°C.

The Raman spectrum (Fig. 2k) contains the typical Lorentzian-shape G-peak at 1598 cm$^{-1}$ and a 2D-peak at 2724 cm$^{-1}$ with a narrow FWHM of 33 cm$^{-1}$ proving the formation of epitaxial monolayer graphene.(Lee et al. 2008) The absence of a distinct D-peak indicates a very low defect density in the hexagonal 2D graphene lattice while the broad D-peak corresponds to the underlying buffer layer (Fromm et al. 2013). This high quality of the graphene layer is another proof for the existence of a well-ordered, initially grown PASG buffer layer because the latter is converted into the graphene layer once the new buffer layer has formed underneath (Emtsev et al. 2008), (Hannon et al. 2011).

The outstanding homogeneity of the PASG graphene is demonstrated by micro-Raman area mappings (each 30 μm x 30 μm) recorded in the center and 500 μm away from the edges of the 5 mm x 10 mm sample. The mappings of the 2D-band FWHM values (Fig. 3c) scatter over a narrow range (blue/green color, left histogram) with a mean value of 33 cm$^{-1}$ and prove that exclusively monolayer graphene was formed (Lee et al. 2008). This was confirmed by large scale optical microscopy inspections (see methods). Thus the sample can be considered as bilayer-free. This is a result of the uniform graphene formation by the PASG and the conservation of low step heights since high step edges are known to favor bilayer formation (Emtsev et al. 2009; Ohta et al. 2010; Kruskopf et al. 2015) (Eriksson et al. 2012). Only at substrate defects such as micro-pipes and in some cases very close to the sample edge bilayer patches were observed (Fig. S1b, supplementary data). Furthermore, the 2D peak positions show a very narrow distribution around 2727 cm$^{-1}$ (Fig. 3c, right histogram) compared to ~2680 cm$^{-1}$ for exfoliated graphene. This indicates a moderate and homogenous compressive strain across the graphene layer, which is a characteristic feature for epitaxial graphene. (Röhrl et al. 2008)

The stabilization of the steps of the SiC surface by PASG is demonstrated by using a substrate with a deliberately larger miscut angle (~0.37°) which usually shows a strong tendency towards giant step bunching at high temperatures. (Virojanadara et al. 2009) This becomes obvious from the AFM image in Fig. 3d (bottom). After standard graphene formation without PASG a completely giant-stepped surface is formed with 1.5 to 2 μm wide terraces and 10 to 15 nm high steps over the entire surface. The corresponding Raman 2D-FWHM mapping (see inset) clearly reveals that monolayer graphene exists on the terraces while along the high terrace edges long stripes of bilayer graphene (FWHM values of 45 - 65 $cm^{-1}$) are present which is typical for graphene step flow growth (Emtsev et al. 2009; Ohta et al. 2010). In contrast, an entirely different and shallow surface structure (Fig. 3d, top) is obtained for a simultaneously processed graphene sample with the pre-defined buffer layer formed by PASG at 1400 °C. Both, the values of the terrace width (approx. ~250 nm) and the heights between 0.5 and 2.5 nm are about 10 times smaller compared to the sample without PASG treatment. Obviously, also in the case of larger miscut angles the increased surface stability achieved by the pre-formed buffer layer successfully suppresses giant step bunching and graphene step flow growth such that a homogenous coverage of monolayer graphene without bilayer stripes is obtained. On such substrates just a few isolated graphene bilayer domains are observed which can be assigned to local step height variations. This experiment demonstrates the high potential of the PASG technique for full-size SiC wafers which usually exhibit fluctuations in their miscut components.

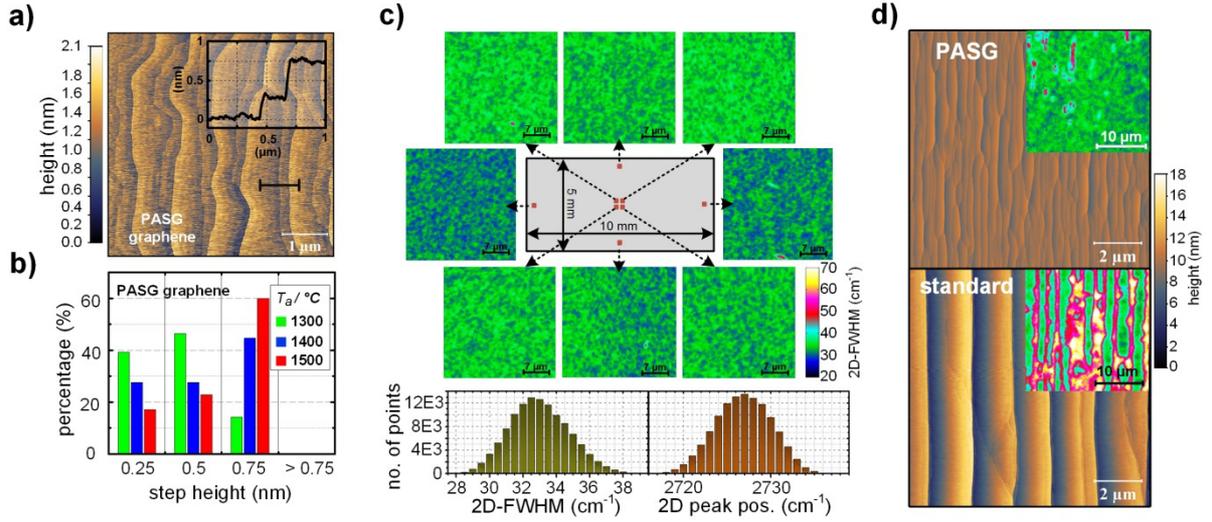

**Figure 3: Morphology of PASG graphene layers. a,** AFM topography (5 μm × 5 μm) of a PASG graphene layer grown at 1750 °C after 5 min annealing at $T_a ≈ 1300$ °C on a 6H-SiC(0001) substrate with a small miscut angle of 0.06°. Inset show the height profile along the indicated line. **b,** Step height distributions of PASG graphene for different pre-annealing temperatures of $T_a$ = 1300, 1400 and 1500 °C. **c,** The Raman mappings (30 μm × 30 μm) of the 2D-peak FWHM were recorded at four positions in the center and another four along the edges (distance to edge = 500 μm): FWHM = 30 - 40 cm$^{-1}$ prove the outstanding homogeneity of the graphene monolayer. Left and right histogram: Distributions of the 2D-FWHM and 2D-peak position calculated at the basis of the eight Raman mappings. **d,** Comparison of PASG with normal sublimation growth of two graphene samples on 6H-SiC(0001) substrates with large miscut angle of 0.37°. Top: AFM topography (10 μm × 10 μm) of the PASG graphene sample with the inset showing a 2D-peak FWHM Raman map (25 μm × 25 μm) with very few isolated bilayer spots of micrometer size. Bottom: AFM image of a simultaneously grown sample without any pretreatment (clean substrate, no hydrogen etching) shows giant step bunching and large bilayer stripes revealed by Raman spectroscopy (FWHM = 45 - 65 cm$^{-1}$, red and yellow shade).

**High-mobility electronic transport properties and quantum metrological device application**

Finally, the electronic properties of the PASG graphene samples (on SiC substrates with small miscut angle) were examined by various transport experiments. Four-probe van der Pauw and Hall measurements were performed on squares of 4.5 mm x 4.5 mm between 2.2 K and 295 K. Measurements performed immediately after finishing graphene growth exhibit an electron mobility of $μ$ = 2800 Vs/cm$^2$ at room temperature and 9500 Vs/cm$^2$ at 2.2 K with an electron density of $n$ = 7.5 × 10$^{11}$ cm$^{-2}$. The measured electron mobilities are comparable and partly better than literature values of state-of-the-art graphene grown by thermal sublimation (Yager et al. 2013) and CVD (Pallecchi et al. 2014). To the best of our knowledge this demonstrates for the first time that high mobilities can be obtained even for very large devices on mm scale in consequence of the excellent homogeneity of the samples.

The measured electron densities below $10^{12}$ cm$^{-2}$ are lower than values usually reported in the literature and close to what is desired for realizing graphene-based resistance standards. In quantum metrology the unit Ohm is traced to the fundamental constants $h$ and $e$ via the von Klitzing constant $R_K \equiv h/e^2$. The main advantage of graphene over a conventional GaAs/GaAlAs based two-dimensional electron system is the larger Landau-level splitting and the absence of a 2$^{nd}$ sub band which makes the quantum Hall effect (QHE) accessible at higher temperatures. In order to examine the quantum Hall resistance plateau at filling factor ν = 2 at reasonable magnetic fields B ≤ 12 T the carrier density was tuned to $n = 1.95 \times 10^{11}$ cm$^{-2}$ by photo-chemical-gating (Lara-Avila et al. 2011). A Hall bar was lithographically processed and aligned nearly perpendicular with respect to the terrace orientation, Fig. 4a. The Hall resistance $R_H$ and the longitudinal resistivity $\rho_{xx}$ as functions of the magnetic flux density $B$ were measured at 4.2 K, Fig. 4b. The high quality of the graphene can be verified by precision measurements of $\rho_{xx}$ in the plateau region since the remaining resistivity in this temperature regime is related to scattering induced by density inhomogeneities (Lafont et al. 2014). For $B > 3$ T a wide resistance plateau of 12.9 kΩ ≈ $R_K$/2 is observed and $\rho_{xx}$ drops below 10 μΩ for $B \geq 10$ T, Fig 4c. $\rho_{xx}$ is thus lower than any published value obtained at 4.2 K (Lafont et al. 2014) (Ribeiro-Palau et al. 2015). This describes extremely low defect scattering and material inhomogeneities despite the perpendicularly orientated terraces with respect to the current direction. Since the 800 step edges along the Hall bar have no negative effect, highly isotropic electronic properties due to the ultra-shallow step heights are expected. From the $\rho_{xx}$ values and the geometry one can calculate an upper limit for the deviation of the Hall resistance measured between any of the three pairs of directly opposing potential contacts (Fig. 4a) from $R_K$/2. At $B = 7$ T one finds $|(R_H - R_K/2)/R_K/2| \lesssim 5 \times 10^{-10}$ which is further reduced to about $1 \times 10^{-10}$ for $B = 10$ T. Note, that previously a comparable performance has only been demonstrated for graphene-based Hall samples at 1.4 K, but not at 4.2 K as in this work (Tzalenchuk et al. 2010) (Lafont et al. 2014) (Ribeiro-Palau et al. 2015). This is an improvement in quantum metrology towards less complex measurement setups and demonstrates the high potential of the PASG technique with respect to graphene-based electronic devices.

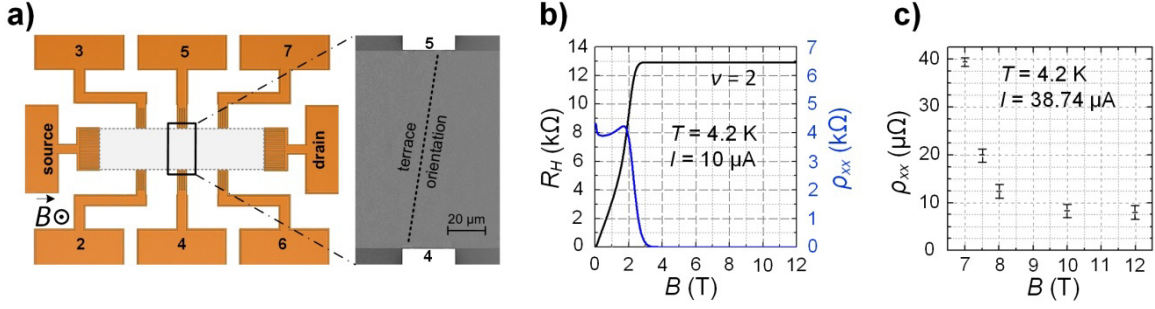

**Figure 4. Magneto-transport measurements of PASG graphene. a,** Left: 100 μm × 400 μm lithographically defined Hall bar (dotted line) with an electron density of $n = 1.95 \times 10^{11}$ cm$^{-2}$ tuned by photo-chemical gating. Right: Optical microscopy in high contrast mode shows the structured bilayer-free graphene region (light grey area) between the contacts 4 and 5. The terrace orientation is orthogonal to the current path. **b,** The quantum Hall resistance (measured between the middle contacts 4 and 5) shows a wide plateau at filling factor $\nu = 2$ with a value of $R_H \approx 12.9$ kΩ corresponding to the expected value of half of the von Klitzing constant $R_K$. Simultaneously, the longitudinal resistivity $\rho_{xx}$ (obtained from measurements involving the outer contacts 3 and 7 at the low-potential side of the Hall bar) approaches zero. **c,** Precision measurements of the longitudinal resistivity in the plateau $\nu = 2$. At $B \geq 10$ T remarkably low $\rho_{xx}$ values below 10 μΩ were measured.

In summary, we have presented a new easy-to-apply graphene fabrication technique in which the sublimation growth is assisted by an extra carbon source from a polymer precursor. The essential step of this mechanism is the stabilization of the SiC surface by enhanced formation of uniformly distributed buffer layer domains from the deposited carbon in the first stage of the thermal process. This is the key to achieve overall suppression of giant step bunching and to enable the use of a larger variety of SiC substrates with different miscut angles. PASG graphene shows bilayer-free graphene monolayer coverage, extraordinarily low terrace step heights below 0.75 nm and excellent electronic properties demonstrated by quantum Hall resistance metrology measurements with an unreached accuracy at 4.2 K. We expect that PASG has the potential to be widely used for large scale graphene-based devices where these material properties are decisive criterions. Considering the large domain size of undisturbed monolayer graphene and the excellent reproducibility, this method represents the last missing part for commercial wafer scale epitaxial graphene growth.

# Methods:

**Substrate and substrate preparation for PASG:** Semi-insulating 6H-SiC(0001) 4" wafer from *II-VI Deutschland* with small nominal miscut angles between 0.01° and 0.06° as well as one with a larger miscut of about 0.37° were used. The wafer was cut into sample pieces of typically 10 mm × 5 mm. For polymer assisted sublimation growth (PASG) a novolac resin is deposited by liquid phase deposition (LPD). Here, we used concentrated AZ5214E photoresist in which the sample was immerged in an ultra-sonic bath for several minutes followed by extensive post-rinsing in isopropanol for about 1 min. The size and density of the droplet-like adsorbate can be used to control the amount of deposited carbon. Alternatively, spin coating of a weak solution of 1 to 4 droplets of AZ-photoresist or PMMA in 50 ml of solvent leads to similar results. (Kruskopf & Pierz 2015), (Momeni Pakdehi 2015)

**Buffer layer and graphene growth:** For buffer layer and graphene growth the samples were introduced into an inductively heated hot-wall reactor. (Ostler et al. 2010) One or two samples were processed at the same time. The sample is put in a graphite suszeptor adjoined by SiC dummy pieces. Initially, the system is evacuated to $1 \times 10^{-6}$ mbar at 950 °C for 30 min. The predefinition of the buffer layer was realized by annealing the sample in an argon atmosphere of 1 bar at temperatures from 1300 to 1500 °C for 5 min. For the subsequent graphene growth the sample was heated (400 K/min) to the growth temperature of 1750 °C and kept for 5 min before cooling down.

The reproducibility of the PASG method is proven by inspection of more than 50 samples by optical microscopy in high contrast reflection imaging mode (Yager et al. 2013), 30 by AFM and 5 by Raman mapping where we have obtained comparable results as presented in this publication.

**AFM:** The Nanostation II AFM from Surface Imaging Systems - SIS (now Bruker) was operated in amplitude-controlled non-contact/intermittent-contact mode. It uses a unique fiber-interferometric detection system that allows absolute measurement of the cantilever tip height. The clear material contrasts in the recorded phase images enables to distinguish SiC, buffer layer and graphene surfaces. The AFM probe type is a *PPP-NCLR* PointProbePlus Silicon sensor from Nanosensors.

**Raman:** Raman measurements were performed using a *LabRAM ARAMIS* spectrometer with an excitation wavelength of 532 nm. A piezo stage enables lateral movement with a step-size of 0.2 μm and a lateral resolution of less than 1 μm. Each single spectrum in Fig. 2k is a difference spectrum based on 121 single measurements distributed over an area of 10 μm × 10 μm to reduce noise of which a similarly recorded spectrum of a clean 6H-SiC reference sample is subtracted to remove disturbing overtones of the substrate. The Micro-Raman mappings in Figure 3c show the distribution of the full-width at half-maximum (FWHM) value of the characteristic 2D peak of graphene (Lee et al. 2008) which was evaluated by an automated Lorentzian fitting algorithm for each data point.

**XPS:** The samples were initially heated to 300°C to remove absorbed contaminations from the sample surface before the measurements were performed in the same UHV environment. The XPS C1s core-level spectra were measured at photon energy of 1486.74 eV which corresponds to the K-alpha line of aluminum. Charging effects caused by the semi-insulating substrates led to a slight shift of the spectra up to a few eV. The uniform peak-shapes indicate a nondispersive shift which allows an alignment of the spectra by shifting the SiC bulk peak to the well-known position at 283.7 eV (Emtsev et al. 2008).

**SPA-LEED:** The high-resolution low energy electron diffraction experiments were performed with a SPA-LEED system. The LEED images were acquired at 140eV and 180eV, which corresponds to a scattering phase of S=4.8 and S=5.4 for the specular spot, respectively.

**Electrical measurements:** Van der Pauw measurements were performed in a He-flow cryostat and a variable magnetic field up to 0.5 T with an automated measurement system. In order to prevent leakage current through graphene on the edges and backside of the sample a square (almost 5 mm × 5 mm) of monolayer graphene on the SiC (0001) face was isolated using a diamond scriber. The graphene was contacted by softly pressing Au contact pins onto the corners of the sample. The ohmic behavior and the linearity of the Hall curves were checked. The samples were measured directly after growth.

The Hall bar (100 μm × 400 μm) for QHE measurements was fabricated using standard e-beam lithography and contacted by Au/Ti bond pads. For the reduction of the electron concentration by chemical gating the graphene sample was spin coated by a PMMA/MMA copolymer of 55 nm followed by a ZEP520A layer of 300 nm. By UV illumination the electron density was adjusted before the sample was cooled down for measurement. The sample was measured at 4.2 K in a helium cryostat equipped with a superconducting magnet.

For the determination of the longitudinal resistivity $\rho_{xx}$ in the QHE measurements with highest resolution a cryogenic current comparator measurement bridge including a 100 Ω normal resistor was used (Goebel et al. 2014). The precision measurements with the results presented in Fig. 4c) followed the established guidelines for the application of the QHE in dc resistance metrology (Delahaye & Jeckelmann 2003).

Additionally, the quantized Hall resistances of the PASG graphene sample was indirectly compared with that of a GaAs-based sample by means of normal resistor with a nominal value equal to ½ RK-90. The difference between both calibration values for the wire-wound resistor was 6 x $10^{-10}$, the expanded relative uncertainty (also including contributions from the corrections to be applied for the normal resistor's air pressure dependence and its drift over six days) was 9 x $10^{-10}$.

# Supplementary Data

**Bi-layer formation on PASG graphene samples at substrate defects and edge regions**

Graphene bi-layer spots after processing at 1750 °C were identified around hexagonal-shaped etch pits that are attributed to crystal defects such as dislocations which randomly occur and locally enhance the decomposition of the SiC, see Figure S1a. In some cases increasing graphene bi-layer content can be observed very close to the sample edge. Figure S1b shows Raman measurements that were recorded only 10 µm away from the sample edge. Usually even here the bi-layer content is negligible and small steps are conserved. The four spectral maps of the 2D-FWHM show that along three sample sides the high quality is conserved, while along one side (area up to 50 to 100 µm away from the edge) giant step bunching (step height of about 5 nm to 15 nm) caused step flow growth and formation of broad bi-layer stripes. Interestingly, this deviant behavior only occurs in the vicinity of a second neighboring SiC sample but not if instead dummy samples (SiC covered by multi-layer graphene) are used. The influence of neighboring samples indicates the high importance of mass transport processes through the gas layer formed by evaporating species above the surface.

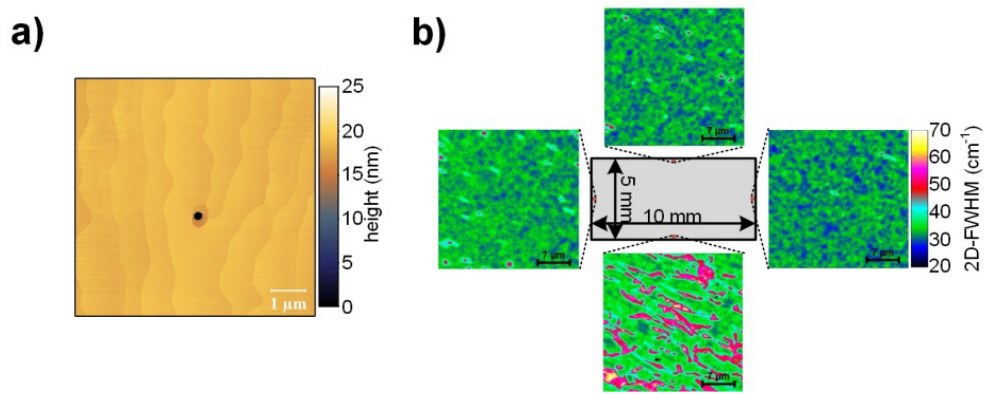

**Figure S1. Reasons for graphene bilayer formation on PASG samples. a,** Substrate defects such as dislocations identified by AFM locally induce formation of deep hexagonal-shape etch pits surrounded by bi-layer graphene. **b,** Usually, even 10 µm close to the sample edge the bilayer content is negligible and small steps are conserved. Only in the vicinity of other neighboring SiC samples instead of multilayer covered dummy samples a direct influence is observed which induces formation of high step edges and bi-layer graphene.

**Raman measurements:**

The D- and G-peak in the Raman spectrum of carbon compounds describe the presence of defects in hexagonal crystal lattices and the presence of and sp² hybridized carbon compounds, respectively. Their intensity ratio I(D)/I(G), peak positions and corresponding FWHM can be used as a measure of carbon network order or disorder. These characteristics allow an approximation of the cluster/crystallite grain size as well as the quantification of chemical bonds with respect to the fraction of sp² and sp³ hybridization. (Tuinstra & Koenig 1970; Ferrari & Robertson 2000)

In the temperature range between 450 °C and 950 °C a transition from amorphous carbon towards nano-crystalline graphite is observed in the Raman spectrum of PASG graphene samples. The broad FWHM of the D- and G-Peak which causes an overlapping of the two components (Figure S2a, top) as well as the large intensity ratio I(D)/I(G) between 0.9 and 1.0 (Figure S2b) prove a high degree of disorder in the material. Additionally, the peak position of the G-peak around ~1587 cm-1 (Figure S2a, bottom) in the discussed temperature range is red-shifted with respect to the typical position of monocrystalline graphite around ~1600 cm-1. The G-peak position and the I(D)/I(G) intensity ratio suggest a fraction of roughly 10% sp³ hybridized carbon for PAG samples annealed at ≥ 450 °C and ≤ 950 °C following the amorphization trajectory by Ferrari and Robertson. The crosslinking of the phenolic resin towards amorphous carbon followed by the transition into nano-crystalline graphite indicates the transition from "stage 2" to "stage 1" in the amorphization trajectory.(Ferrari & Robertson 2000) A very similar behavior was observed by Ko et al. who investigated the pyrolysis of phenolic resin in a wide temperature range (Ko et al. 2000).

The crystallite size of carbon nano-crystals can be estimated from the Tuinstra-Koenig relation and the I(D)/I(G) intensity ratio in the Raman spectrum. The model was applied to samples with amorphous carbon and nano-crystalline graphite and to those which describe the very first formation of buffer layer. The corresponding crystallite sizes are approximately 4-5 nm after annealing at 950 °C with an upwards trend for higher temperatures. (Tuinstra & Koenig 1970) However, the Tuinstra-Koenig relation may not be valid in the presence of buffer layer or graphene domains content due to the increasing interactions of the carbon layers and the SiC substrate which causes changes in the shape and positions of the D- and G-peaks. This can be understood from the samples annealed at 1400 °C and higher. Here the D- and G-peak clearly shifts to larger wavenumbers which is not conform to the amorphization trajectory model for increasing structural order in the carbon network towards nano-crystalline graphite. The narrowing of the D- and G-peak, the rise of new Raman modes around 1490 cm-1, the overall reduced intensity and the absence of the 2D-peak of graphene are characteristics that describe the formation of the buffer layer. The interaction of this carbon layer with the SiC substrate leads to differences in the phonon dispersion and the electronic states compared to free-standing graphene or graphite as described by Fromm et al. (Fromm et al. 2013). This would explain the different between the Raman spectra of samples annealed at ≥ 1400 °C and those of lower annealing temperatures were no buffer layer has formed. After graphene formation at 1750 °C a sharp G-peak typically appears at 1598 cm$^{-1}$ along with the 2D-peak at 2727 cm$^{-1}$. No sharp graphene related D-Peak at ~1362 cm$^{-1}$ occurs which indicates a low defect density of PASG graphene. The remaining broad background from 1300 cm$^{-1}$ to 1600 cm$^{-1}$ overlapping the typical sharp G-peak in the graphene Raman spectrum most probably correspond to the still visible Raman modes of the underling buffer layer as also proposed by Fromm et al. (Fromm et al. 2013)

In addition to Raman spectroscopy investigations the same spectrometer was used to analyze photoluminescence of the PASG samples in relation to the annealing temperature (Figure S2c). Especially for the sample annealed at 450 °C a clear photoluminescence band (PL) was observed in a spectral range between 600 nm and 800 nm which superimposes the Raman spectra. After high temperature annealing at ≥ 1400°C the effect clearly disappears. Photoluminescence in amorphous carbon was already mentioned by Casiraghi et al., explaining this effect as a result of hydrogenated carbon chains in amorphous carbon. The intensity ratio of the PL-band with respect to the G-peak is a measure for the hydrogen content. Therefore, we expect a certain amount of C-H bonds up to temperatures of ≤ 1300 °C (Casiraghi et al. 2005).

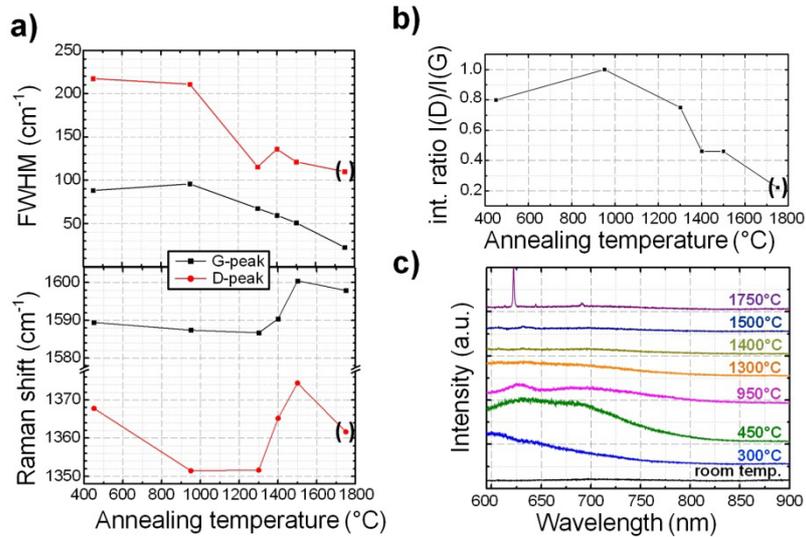

**Figure S2. Temperature dependent characteristics in the Raman spectrum of PASG samples. a,** The conversion of the polymer by annealing towards amorphous carbon, nano-crystalline graphite, buffer layer and graphene is accompanied by changes in the FWHM and positions of the D- and G-peak. The increasing structural order during buffer layer and graphene formation is visualized by the narrowing of the FWHM (top) of the D- and G-Peak from 950 °C to 1750 °C and a shift of the peak positions (bottom) towards higher wavenumbers. **b,** Simultaneously, the intensity ratio I(D)/I(G) significantly drops from ~1 towards ~0.2. Data points noted in brackets involve the D-Peak of the buffer layer that overlaps the spectrum of the graphene layer. **c,** A clear photoluminescence band was observed for the sample annealed at ≥ 450 °C and ≤ 1300 °C which is expected to be due to remaining hydrogen-carbon bonds originating from the polymer (Casiraghi et al. 2005).

# Acknowledgements

We gratefully acknowledge funding by the *School for Contacts in Nanosystems (NTH)* and the support by the *Braunschweig International Graduate School of Metrology B-IGSM* and *NanoMet*.

# Author contributions

K.P. and M.K conceived the idea and wrote the manuscript and performed characterization analysis, M.K. performed the experiments including synthesis of material and device fabrication, D.M.P. improved the PASG technique, S.W. and R.S. performed detailed sample characterization by Raman spectroscopy, T.Z. and M.K. characterized the samples by AFM, M.G. performed precision QHE measurements, J.B., J.A. and C.T. characterized the samples by SPA-LEED, J.L. and T.S. characterized the samples by LEED and XPS, All authors discussed, read and commented on the manuscript.